\documentclass[10pt, conference, letterpaper]{IEEEtran}
\usepackage{cite}
\usepackage{amsmath,amssymb,amsfonts}
\usepackage{algorithmic}
\usepackage{graphicx}
\usepackage{textcomp}
\usepackage{xcolor}
\usepackage{listings}
\usepackage{booktabs}
\usepackage{tikz}
\usepackage{url}
\usepackage{booktabs}
\usepackage{array}
\usepackage{caption}
\usepackage{multicol}
\usepackage{enumitem}
\usetikzlibrary{shapes.geometric, arrows.meta, positioning, calc}

\setlist[itemize]{noitemsep, topsep=2pt, parsep=0pt, partopsep=0pt}
\setlist[enumerate]{noitemsep, topsep=2pt, parsep=0pt, partopsep=0pt}
\setlist[description]{noitemsep, topsep=2pt, parsep=0pt, partopsep=0pt}
\definecolor{keywordcolor}{rgb}{0.1, 0.1, 0.6}
\definecolor{prefixcolor}{rgb}{0.5, 0.0, 0.5}
\definecolor{stringcolor}{rgb}{0.0, 0.5, 0.0}
\definecolor{commentcolor}{rgb}{0.4, 0.4, 0.4}

\lstdefinelanguage{turtle}{
  morekeywords={a},
  sensitive=true,
  morecomment=[l]{\#},
  morestring=[b]",
  alsodigit={-},
  classoffset=0,
  keywordstyle=\color{keywordcolor}\bfseries,
  classoffset=1,
  morekeywords={@prefix, @base},
  keywordstyle=\color{prefixcolor}\bfseries,
  classoffset=0
}

\tikzset{
    icm/.style={
        rectangle, draw=black, double, fill=gray!12, text=black, 
        font=\rmfamily\bfseries\scriptsize, minimum width=2.4cm, minimum height=0.5cm, align=center
    },
    sec/.style={
        rectangle, draw=black, fill=white, thick, text=black, 
        font=\rmfamily\scriptsize, minimum width=3.0cm, minimum height=0.5cm, align=center
    },
    ext/.style={
        rectangle, draw=black, dashed, fill=gray!5, text=black, 
        font=\rmfamily\scriptsize, minimum width=2.2cm, minimum height=0.5cm, align=center
    },
    edgeLabel/.style={
        font=\rmfamily\tiny, inner sep=1.5pt, fill=white, opacity=0.95, text=black
    },
    blackArrow/.style={
        -{Stealth[scale=0.75]}, thick
    },
    blueArrow/.style={
        -{Stealth[scale=0.75]}, thick, dash pattern=on 3pt off 2pt, draw=black!70
    }
}

\begin{document}

\title{A Standardized Ontology for Intent-Based Security Management in Autonomous Networks}

\author{
    \IEEEauthorblockN{Loay Abdelrazek}
    \IEEEauthorblockA{
        AI \& Technology Foresight \\
        Ericsson, Sweden \\
        first.last@ericsson.com
    } \and
     \IEEEauthorblockN{Sultan Ertas}
    \IEEEauthorblockA{
        Ericsson Research Networks \\
        Ericsson, Türkiye \\
        first.last@ericsson.com
    } \and
     \IEEEauthorblockN{Andrey Silva}
    \IEEEauthorblockA{
        Ericsson Research Networks \\
        Ericsson, Brazil \\
        first.last@ericsson.com
    } 
}

\maketitle

\begin{abstract}
Modern 5G-Advanced and emerging 6G architectures face complex, multi-layered threat vectors that outpace traditional manual security configurations. Shifting security management towards autonomous, self-protecting operation requires formal semantic frameworks. This work specifies the TM Forum TR292I Security Ontology v4.0.0, a standardized Resource Description Framework Schema (RDFS) compliant vocabulary for declarative security management. By natively extending the TM Forum Intent Common Model (ICM), the ontology decouples high-level security goals from underlying technical controls. Crucially, it embeds resource cost mapping properties to ensure autonomous mitigation actions safeguard Service Level Agreements (SLAs). We validate this model-driven architecture through a formal semantic walkthrough of a distributed Denial of Service (DDoS) mitigation sequence on a disaggregated Next-Generation NodeB (gNB) slice using W3C Turtle and SPARQL. The results demonstrate that runtime constraint conflicts are resolved dynamically without human intervention, establishing a reproducible framework for standardized, intent-driven network security orchestration.
\end{abstract}


\section{Introduction}
Mobile network management is undergoing a paradigm shift. As operators deploy cloud-native 5G-Advanced and emerging 6G architectures, legacy operational models fail against escalating complexity\cite{zeydan2025network}. Next-generation networks are no longer monolithic pipe structures; they are fluid, highly distributed webs of containerized network functions, dynamic network slices, edge computing resources, and diverse multi-vendor ecosystems.

Concurrently, threat actors leverage sophisticated, multi-stage campaigns targeting multiple computing stack layers simultaneously \cite{oran_sec}. From radio interface jamming and signaling storms to sophisticated control-plane exploits and API supply-chain vulnerabilities, the entry points are shifting faster than traditional security teams can track.

Traditional manual approaches to configuring security controls (encompassing both preventive mechanisms such as firewalls and detection capabilities such as jamming or radio signaling storm detection) are no longer adequate to address the pace and sophistication of emerging threats \cite{enisa5g2022}. When human engineers are required to manually analyze security incidents, update controls and detection capabilities, determine the appropriate security posture, and deploy configurations across thousands of cell sites, the response latency introduced by such processes renders the network vulnerable before remediation can be completed \cite{batewela2025addressing}.

The limitations of traditional, rule-based operations are most visible in modern disaggregated architectures like Open Radio Access Networks (O-RAN). In an O-RAN deployment, a single logical gNB is broken down into separate physical or virtual components: the Open Central Unit (O-CU), Open Distributed Unit (O-DU), and Open Radio Unit (O-RU), interconnected over dynamic xHaul interfaces. Traditional security enforcement relies on rigid, imperative logic: \textit{"If Condition A occurs, execute Action B on Appliance C."} This approach breaks down completely within a software-defined, virtualized ecosystem \cite{leivadeas2022survey}. If an O-CU plane automatically scales out or migrates to a different cloud edge node, its topology and IP structures shift instantly. Under these dynamic conditions, static, imperative rules become a liability rendering security enforcement blind to the changes or inadvertently creating self-inflicted Denial of Service (DoS) routing bottlenecks that sever critical control loops.

To achieve resilient, autonomous operations, the telecom industry must pivot to a self-protecting infrastructure driven by automated closed-loop architectures. To bridge this structural gap, we designed and introduced the TM Forum TR292I Security Ontology v4.0.0 \cite{tmf_tr292i} as part of the Autonomous Networks project team. This specification establishes the formal, mathematical, and structural vocabulary needed to turn intent-based operations into reality.

Intent-based networking represents a shift from imperative configuration to declarative state management. Rather than prescribing technical procedures step-by-step, an operator specifies the desired high-level operational state as an intent. The network infrastructure, through an intelligent component referred to as the Intent Management Function (IMF), interprets this intent, evaluates the real-time operational context of the network, and derives the necessary configuration changes. When deviations from the intended state are detected, the underlying closed-loop automation mechanism identifies the discrepancy and autonomously determines the appropriate remediation actions without requiring human intervention.

Within the IMF sits our TR292I v4.0.0 ontology, which provides explicit RDFS vocabularies that inherit from and extend the foundational TM Forum ICM. This architectural inheritance ensures that heterogeneous multi-vendor security systems, network orchestrators, and agentic reasoning components share a unified structural representation of threats, capabilities, and cross-domain system impacts.

The primary contributions of this paper are summarized as follows:
\begin{itemize}
    \item \textbf{Standardized Security Intent Model Design:} We present the architectural design principles behind our standardized contribution, the TM Forum TR292I v4.0.0 ontology, detailing how it decouples functional security expectations from resource-consuming physical countermeasures.
    \item \textbf{Multi-Constraint Optimization Formulation:} We formalize the logical engine utilized by the IMF into a multi-constraint optimization problem, explicitly defining how security intent compliance is maintained without violating active consumer service level agreements (SLAs).
    \item \textbf{Formal Semantic Verification Walkthrough:} We validate the structural integrity of our model-driven architecture through a real-time DDoS mitigation sequence on a virtualized gNB, demonstrating semantic candidate evaluation using explicit SPARQL and Description Logic graph processing.
\end{itemize}

\section{Standards Gap Analysis}
The realization of autonomous network management has catalyzed structural standardization efforts across multiple standardization bodies. This section reviews the most relevant existing standards and frameworks, identifies their architectural limitations with respect to intent-based security management, and positions the contributions of the TR292I v4.0.0 ontology in relation to these identified gaps.

\subsection{IETF Intent-Based Networking (IbN)}
The IETF framework (codified primarily across RFC 9315 \cite{ietf_rfc9315} and emerging Network Management Datastore Architecture drafts) focuses deeply on resource slice carving, topological mapping, and transport provisioning. It excels at establishing virtual private networks (VPNs) and slicing tunnels based on abstract traffic profiles. However, IETF IbN lacks built-in semantic models to decouple security vectors from general operational noise. Within IETF IbN, security parameters are typically treated as general policy attributes or static metadata tags embedded within a standard Yet Another Next Generation (YANG) data model, preventing the system from deducing real-time risk relationships.

\subsection{ETSI Zero-Touch Management (ZSM)}
ETSI ZSM \cite{etsi_zsm} introduces highly reliable, decoupled closed execution loops (Monitor, Analyze, Plan, Execute and Knowledge - MAPE-K) designed for end-to-end telecom architecture softwarization. While ZSM explicitly accounts for cross-domain security monitoring, it enforces imperative, structural policy bindings that create technical friction across multi-vendor boundaries. The framework handles events sequentially via rigid threshold rules; it lacks a standardized, decoupled knowledge graph capable of describing how an isolated security threat correlates directly to shared physical or logical components. Furthermore, ZSM execution loops do not natively evaluate whether a selected mitigation action will introduce catastrophic SLA drops on the very infrastructure planes they are tasked to protect.

\subsection{ETSI Experiential Networked Intelligence (ENI)}
The ETSI ENI specification \cite{etsi_eni} defines an AI-driven cognitive network management architecture that uses semantic processing and policy-driven contexts to adjust network configurations dynamically. ENI relies on an internal system knowledge graph to capture the operating context of the infrastructure. Despite its inclusion of context-aware ontologies, ETSI ENI introduces a profound architectural limitation regarding security management: it fails to establish a decoupled taxonomy for security abstractions. In an ENI-driven system, functional properties and non-functional security goals are processed within a singular, massive reasoning pipeline, causing computational bottlenecks during active network attacks. Furthermore, ENI does not provide a standardized class schema to model the quantitative resource degradation costs introduced by its own cognitive countermeasures.

\subsection{The Architectural Contribution of TR292I v4.0.0}
The TR292I v4.0.0 ontology introduces a comprehensive semantic domain vocabulary dedicated to expressing intent-based, declarative security expectations within autonomous networks. Prior to this specification, intent frameworks could express general network configurations (such as slicing or traffic routing profiles), but lacked the structural vocabulary required to define abstract security constraints, map vulnerability boundaries, or orchestrate multi-vendor countermeasures.

TR292I bridges this critical gap through two core architectural design patterns. First, it directly extends the foundational TM Forum Intent Common Model (ICM), ensuring that security expectations inherit native intent structures seamlessly. Second, it decouples the high-level, functional security goals from technical enforcement capabilities while concurrently mapping deployment costs via the \texttt{sec:impactOnResource} property. By keeping security intents syntactically differentiatied from general network operations and building upon RDFS hierarchies, our architecture allows specialized security reasoners to process security states within short time windows while preserving end-to-end SLA optimization.


\begin{table*}[t]
\caption{Standards Comparison Matrix}
\label{tab:standards_comparison}
\centering
\small
\renewcommand{\arraystretch}{1.2} 
\begin{tabular}{l p{3.5cm} p{3.5cm} p{3.5cm}}
\toprule
\textbf{Architectural Metric} & \textbf{IETF IbN} & \textbf{ETSI ZSM} & \textbf{\textbf{TM Forum TR292I v4.0.0}} \\
\midrule
\textit{Primary Abstraction} & Network Slices / VPN Tunnels & Closed-Loop Policy Integration & \textbf{Formalized Security Intent Domain Vocabulary} \\
\addlinespace
\textit{Security Modeling} & Implicit (General attributes) & Imperative Rules / Policies & \textbf{Declarative Classes Extended via TMF ICM} \\
\addlinespace
\textit{Implementation Layer} & Bound to Resource YANG Models & Coupled to Orchestrator Scripts & \textbf{Decoupled (Expectation vs. Capability Classes)} \\
\addlinespace
\textit{Resource Impact Aware} & Static Profile Constraints & External Telemetry Hooks & \textbf{Native Cost Mapping} \\
\addlinespace
\textit{Conflict Resolution} & Manual Rule Engineering & Rule-Based Orchestration & \textbf{Model-Driven Automated Graph Reasoning} \\
\bottomrule
\end{tabular}
\end{table*}

\section{Architectural Principles and Design of the TR292I Model}
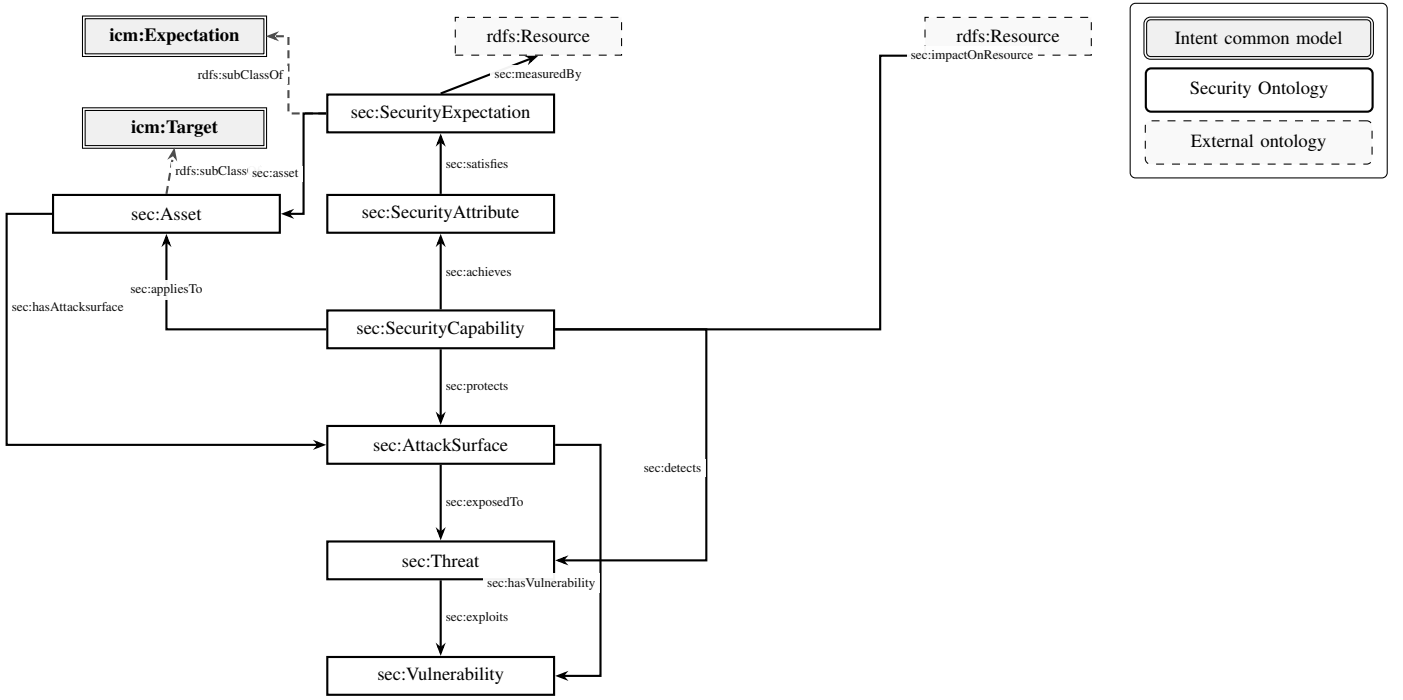
\begin{figure*}[!t]
\centering
\begin{tikzpicture}[node distance=1.2cm and 2.2cm]

    \node[icm] (icmExp) {icm:Expectation};
    \node[ext, right=2.5cm of icmExp] (rdfsRes1) {rdfs:Resource};
    \node[ext, right=4.0cm of rdfsRes1] (rdfsRes2) {rdfs:Resource};

    \node[icm, below=0.7cm of icmExp] (icmTarget) {icm:Target};
    \node[sec, below right=0.5cm and 0.8cm of icmExp] (secExp) {sec:SecurityExpectation};

    \node[sec, below left=0.8cm and 0.6cm of secExp] (secAsset) {sec:Asset};
    \node[sec, below=0.8cm of secExp] (secAttr) {sec:SecurityAttribute};

    \node[sec, below=1.0cm of secAttr] (secCap) {sec:SecurityCapability};

    \node[sec, below=1.0cm of secCap] (secAttack) {sec:AttackSurface};

    \node[sec, below=1.0cm of secAttack] (secThreat) {sec:Threat};

    \node[sec, below=1.0cm of secThreat] (secVuln) {sec:Vulnerability};
    
    \draw[blueArrow] (secExp.west) -- ++(-0.5,0) |- (icmExp.east) 
        node[edgeLabel, pos=0.25, left] {rdfs:subClassOf};
    \draw[blueArrow] (secAsset.north) -- (icmTarget.south) 
        node[edgeLabel, midway, right] {rdfs:subClassOf};

    \draw[blackArrow] (secExp.north) -- (rdfsRes1.south) 
        node[edgeLabel, midway, right] {sec:measuredBy};
    \draw[blackArrow] (secCap.east) -- ++(4.3,0) |- (rdfsRes2.south) 
        node[edgeLabel, pos=0.6, right] {sec:impactOnResource};

    \draw[blackArrow] (secAttr.north) -- (secExp.south) 
        node[edgeLabel, midway, right] {sec:satisfies};
    \draw[blackArrow] (secCap.north) -- (secAttr.south) 
        node[edgeLabel, midway, right] {sec:achieves};
    \draw[blackArrow] (secCap.south) -- (secAttack.north) 
        node[edgeLabel, midway, right] {sec:protects};
    \draw[blackArrow] (secAttack.south) -- (secThreat.north) 
        node[edgeLabel, midway, right] {sec:exposedTo};
    \draw[blackArrow] (secThreat.south) -- (secVuln.north) 
        node[edgeLabel, midway, right] {sec:exploits};

    \draw[blackArrow] (secExp.west) -- ++(-0.3,0) |- (secAsset.east) 
        node[edgeLabel, pos=0.3, left] {sec:asset};
    \draw[blackArrow] (secCap.west) -- (secAsset.south |- secCap) -- (secAsset.south) 
        node[edgeLabel, pos=0.3, above] {sec:appliesTo};
    \draw[blackArrow] (secAsset.west) -- ++(-0.6,0) |- (secAttack.west) 
        node[edgeLabel, pos=0.2, right] {sec:hasAttacksurface};

    \draw[blackArrow] (secCap.east) -- ++(2.0,0) |- (secThreat.east) 
        node[edgeLabel, pos=0.3, left] {sec:detects};
    \draw[blackArrow] (secAttack.east) -- ++(0.6,0) |- (secVuln.east) 
        node[edgeLabel, pos=0.3, left] {sec:hasVulnerability};

    \matrix [draw=black, thin, fill=white, rounded corners=2pt, below right=0.1cm and 0.5cm of rdfsRes2, yshift=0.8cm, inner sep=5pt] {
        \node[icm, minimum width=3cm, minimum height=0.4cm, font=\rmfamily\scriptsize] {Intent common model}; \\
        \fill[white] (0,0) rectangle (0,0.1); \\ 
        \node[sec, minimum width=3cm, minimum height=0.4cm, font=\rmfamily\scriptsize] {Security Ontology}; \\
        \fill[white] (0,0) rectangle (0,0.1); \\ 
        \node[ext, minimum width=3cm, minimum height=0.4cm, font=\rmfamily\scriptsize] {External ontology}; \\
    };
\end{tikzpicture}
\caption{The proposed security ontology mapping architecture.}
\label{fig:ontology_model}
\end{figure*}

As illustrated in Fig. \ref{fig:ontology_model}, the TR292I ontology establishes an interconnected relationship graph between the most crucial concepts for security management. In this fourth version of the ontology, the model has been streamlined, stripping away heavy structural dependencies to focus intensely on core intent classes. TR292I is architected around RDFS vocabularies. This schema leverages clean structural hierarchies and structural inheritance from the foundational TM Forum ICM \cite{tmf_tr290} to minimize processing footprints and inherit similar vocabularies and avoid unnesscary repition of concepts.

    
    

    


The following subsections dissect the structural definitions, class properties, and relationships of the core TR292I vocabulary. Specifically, we demonstrate how these abstract declarations map operational security goals directly to dynamic infrastructure instances, organizing the ontology into its primary building blocks.

\subsection{Dynamic Asset Declaration for Adaptive Security Management}
Within the TR292I specification, a \texttt{sec:Asset} represents any high-value logical, software-defined, or physical resource within the telecommunications stack that possesses explicit security requirements. A critical design decision in version 4.0.0 is the formalization of \texttt{sec:Asset} as a direct \texttt{rdfs:subClassOf} of \texttt{icm:Target} from the foundational TM Forum ICM. Because an asset inherits core intent target properties, it processes complex resources structurally as an \texttt{rdfs:Container}. This architectural choice provides exceptional runtime flexibility. Instead of hardcoding unique, static network identifier strings into rigid firewall tables, an operator can declare assets dynamically using contextual criteria, metadata flags, and operational state filtering.

\subsection{Formalization of Attack Surface for Vulnerability Awareness}
Exposure vectors and perimeter boundaries are explicitly captured via the \texttt{sec:AttackSurface} class. Modeled structurally as a subclass of \texttt{rdfs:Container}, an attack surface represents either an isolated network interface (e.g., an O-RAN N2/F1 control plane interface) or groups thousands of distributed entry points that share identical security vulnerabilities. By establishing an explicit link between a \texttt{sec:Asset} and a \texttt{sec:AttackSurface} via the \texttt{sec:attackSurface} relationship property, the ontology gives the IMF automated structural awareness of exactly where an asset is vulnerable to potential compromise.

\subsection{Formalization of Security Expectations for Intent Expressions}
The core declarative security goals of an intent expression are categorized using the \texttt{sec:SecurityExpectation} class, which serves as a specialized subclass of \texttt{icm:Expectation}. As illustrated in Fig. \ref{fig:ontology_model}, TR292I splits this abstraction into two operational classes to separate functional protective state conditions from non-functional, telemetry-driven validation loops:
\begin{itemize}
    \item \textbf{Protection Expectations \\(\texttt{sec:ProtectionExpectation}):} Dictate the specific abstract security goals (e.g., \texttt{sec:Confidentiality}, \texttt{sec:Integrity}, or \texttt{sec:Availability}) that must be enforced across an asset's attack surface using the \texttt{sec:satisfies} property relation.
    \item \textbf{Performance Expectations \\(\texttt{sec:PerformanceExpectation}):} Govern the non-functional, quantifiable, and bound parameters of security enforcement, leveraging the \texttt{sec:measuredBy} property to map security expectations directly to live telemetry streams and quantitative performance indicators.
\end{itemize}

\subsection{Security Capability Modeling Across the Execution Plane}
The technical countermeasures, software controls, and cryptographic primitives deployed across the infrastructure are modeled via the \texttt{sec:SecurityCapability} class. TR292I organizes these execution-plane components into three distinct operational subclasses:
\begin{enumerate}
    \item \texttt{sec:PreventiveCapability}: Proactive infrastructure hardening mechanisms, such as IPsec tunnel enforcement or TLS mutual authentication.
    \item \texttt{sec:DetectionCapability}: Observation and passive packet scanning tools, such as Deep Packet Inspection (DPI) probes and signaling anomaly sensors.
    \item \texttt{sec:MitigationCapability}: Dynamic, reactive network actions designed to neutralize active threat states, such as automated request throttling or flow re-routing.
\end{enumerate}

\subsection{The Multi-Constraint Optimization Engine}
A primary architectural challenge in autonomous security loops is resolving cross-domain operational conflict. When a security mitigation capability is activated, it inevitably consumes underlying infrastructure resources, which can degrade consumer application performance. TR292I addresses this challenge natively by modeling security capabilities as consumers of active network infrastructure. Using the properties \texttt{sec:impactOnResource} and \texttt{sec:impactValue}, a capability announces its structural cost profile (such as transport latency penalties or CPU core consumption) to the IMF knowledge graph before enforcement.

Formally, let $C$ be the set of available candidate mitigation capabilities discovered within the local schema repository ($c \in C$). Let $P$ represent the maximum acceptable security processing time constraint declared within the performance expectation, and let $R$ be the set of critical operational resource types monitored under active consumer SLAs. The IMF resolves candidate selection by solving the following multi-constraint optimization problem:
\begin{equation}
    \text{Select } c^* = \arg\min_{c \in C} \left( \mathcal{M}_{\text{time}}(c) \right)
\end{equation}
\begin{equation}
    \text{subject to: } \mathcal{M}_{\text{time}}(c) \leq P_{\text{max\_mitigation\_time}}
\end{equation}
\begin{equation}
    \forall r \in R, \quad \mathcal{I}_{\text{value}}(c, r) \leq \mathcal{V}_{\text{ceiling}}(r)
\end{equation}
where $\mathcal{M}_{\text{time}}(c)$ represents the performance metric value defining the capability's response speed, $\mathcal{I}_{\text{value}}(c, r)$ is the native resource impact cost introduced by capability $c$ on resource type $r$, and $\mathcal{V}_{\text{ceiling}}(r)$ represents the absolute performance degradation threshold permitted by the active service tier. By embedding this constraint model directly within the graph architecture, the system prevents autonomous security actions from inadvertently violating broader operational metrics.

\section{Analytical Evaluation and Case Study Walkthrough}
To demonstrate the mathematical and structural validity of the proposed meta-model without relying on an idiosyncratic physical testbed, this section presents a formal semantic execution walkthrough using standard W3C Turtle syntax \cite{w3c_turtle}.

\subsection{Scenario Setup}
We consider an operational 5G network slice supporting an array of automated industrial drones. The primary network asset is a Next-Generation NodeB cell interface container (\texttt{ex:gNB01}) targeted by an active malicious control-plane signaling attack (\texttt{ex:RadioDdosThreat}). The operator has submitted an intent specifying two conflicting expectations: (1) Any active threat must be neutralized within 50 ms (\texttt{ex:MeanTimeToMitigate}). (2) The countermeasure must not introduce more than a 20ms penalty to the end-to-end network latency metric (\texttt{ex:NetLatency}).

To evaluate how our security ontology models these exact real-time requirements, the operational layout is formalized using the Terse RDF Triple Language (Turtle) serialization format.

\subsection{Intent Initialization and Modelling Assets}
The evaluation framework relies on common structures from the TM Forum Intent Common Model (ICM) alongside our custom security ontology layer. Listing~\ref{lst:prefixes} outlines the initialization of the intent object, assets and attack surfaces.



\begin{lstlisting}[language=turtle, caption={Modelling Example usng TR292I Security Ontology: Intent and Assets\vspace{5pt}}, label={lst:prefixes}]
# Intent
ex:DdosMitigationIntent
    a icm:Intent ;
    rdfs:label "DDoS mitigation intent for a disaggregated gNB slice" ;
    log:allOf (
        ex:DdosSecurityPropertyExpectation
    ) .

# Target
ex:gNB01Target
    a icm:Target ;
    rdfs:member ex:gNB01 .

ex:gNB01
    a sec:Asset ;
    rdfs:label "Next-Generation NodeB Slice Asset" ;
    sec:attackSurface ex:ControlPlaneSurface .

ex:ControlPlaneSurface
    a sec:AttackSurface ;
    rdfs:member ex:CP_N2_Interface .

ex:CP_N2_Interface
    a sec:NetworkInterface ;
    rdfs:label "N2 Control Plane Interface" .
\end{lstlisting}

By tracking \texttt{ex:gNB01} as a member of \texttt{icm:Target}, the intent handler identifies the asset boundary. The relationship predicate \texttt{sec:attackSurface} links the parent container to the underlying \texttt{ex:CP\_N2\_Interface}, which maps directly to the targeted control-plane interface mentioned in the scenario and identified as the attack surface to be protected.

\subsection{Modelling Performance Property Conditions}
To resolve the conflicting performance envelopes requested by the operator, the intent translates qualitative metrics into strict quantitative checks. Listing~\ref{lst:conditions} defines the semantic conditions using the TM Forum quantity model.

\begin{lstlisting}[language=turtle, caption={Modelling Example usng TR292I Security Ontology: Expectations and Conditions\vspace{5pt}}, label={lst:conditions}]

# Property Expectation
ex:DdosSecurityPropertyExpectation
    a icm:PropertyExpectation ;
    icm:target ex:gNB01Target ;
    rdfs:label "DDoS mitigation performance and latency property expectation" ;
    log:allOf (
        ex:TargetMitigationTimeCondition
        ex:MaxLatencyCeilingCondition
    ) .

# Condition 1: mitigation time must be at most 50 ms
ex:TargetMitigationTimeCondition
    a log:Condition ;
    rdfs:label "Mean time to mitigate shall be at most 50 ms" ;
    quan:atMost (
        ex:MeanTimeToMitigate
        [
            rdf:value "50"^^xsd:integer ;
            quan:unit "ms"
        ]
    ) .

# Condition 2: latency impact must be at most 20 ms
ex:MaxLatencyCeilingCondition
    a log:Condition ;
    rdfs:label "Net latency impact shall be at most 20 ms" ;
    quan:atMost (
        ex:NetLatencyImpact
        [
            rdf:value "20"^^xsd:integer ;
            quan:unit "ms"
        ]
    ) .
\end{lstlisting}
The logical \texttt{log:allOf} operator mandates that a candidate action must satisfy both structural bounds. The thresholds match the scenario specifications exactly: the time-to-mitigate ceiling is set to 50~ms via \texttt{ex:MeanTimeToMitigate}, and the latency penalty is capped at 20~ms via \texttt{ex:NetLatencyImpact}.

\subsection{Mitigation Candidate Evaluation and Trade-offs}
The physical infrastructure exposes different enforcement capabilities with varied performance trade-offs. Listing~\ref{lst:dataset_capabilities} lists the profiles for the competing candidate capabilities.

\begin{lstlisting}[language=turtle, caption={Modelling Example usng TR292I Security Ontology: Mitigation Capabilities\vspace{5pt}}, label={lst:dataset_capabilities}, float]
# Candidate Capability A: DPI Redirect
ex:DpiCapability
    a sec:MitigationCapability ;
    rdfs:label "DPI Redirect" ;
    sec:mitigates ex:RadioDdosThreat ;
    sec:measuredBy  [ ex:MeanTimeToMitigate ;
        rdf:value "10"^^xsd:integer ;
        quan:unit "ms"
    ] ;
    sec:impactOnResource [ ex:NetLatencyImpact ;
    rdf:value "35"^^xsd:integer ;
    quan:unit "ms"
    ] .

# Candidate Capability B: Edge Throttling
ex:ThrottlingCapability
    a sec:MitigationCapability ;
    rdfs:label "Edge Throttling" ;
    sec:mitigates ex:RadioDdosThreat ;
    sec:measuredBy [ ex:MeanTimeToMitigate ;
    rdf:value "30"^^xsd:integer ;
    quan:unit "ms"
    ] ;

    sec:impactOnResource [ ex:NetLatencyImpact ;
    rdf:value "15"^^xsd:integer ;
    quan:unit "ms"
    ] .

\end{lstlisting}
When the policy loop processes the scenario requirements against these candidates, it identifies a clear trade-off:
\begin{itemize}
    \item \textbf{\texttt{ex:DpiCapability}}: Neutralizes the threat extremely fast (10~ms), but fails the latent constraint by introducing a heavy 35~ms routing penalty.
    \item \textbf{\texttt{ex:ThrottlingCapability}}: Mitigates the threat slower (30~ms), but stays under the latency penalty ceiling at 15~ms. 
\end{itemize}

Consequently, the ontology engine selects the edge throttling capability as the only fully compliant enforcement plan.

\subsection{Agentic Closed-Loop Execution Sequence}
When the attack occurs, the closed-loop system steps through the following autonomous validation sequence:
\begin{enumerate}
    \item \textbf{Detection:} An edge anomaly sensor (\texttt{sec:DetectionCapability}) identifies a signaling anomaly and asserts that \texttt{ex:RadioDdosThreat} is active against \texttt{ex:gNB01}.
    \item \textbf{Query \& Evaluation:} The IMF engine queries the operational graph database using a standardized semantic protocol. To discover valid countermeasures dynamically, the agent executes the declarative SPARQL query shown in Listing \ref{lst:sparql}, returning candidates and their native resource costs shown in Listing 2.
\end{enumerate}

\begin{figure}[t]
\begin{lstlisting}[language=SQL, caption={SPARQL Countermeasure Discovery Query\vspace{5pt}}, label={lst:sparql}, basicstyle=\ttfamily\tiny, frame=single, backgroundcolor=\color{white}, keywordstyle=\bfseries]
SELECT ?capability ?latencyCost ?mitigateTime
WHERE {
  ?capability a sec:MitigationCapability ;
              sec:mitigates ex:RadioDdosThreat ;
              sec:impactOnResource ex:NetLatencyImpact ;
              sec:impactValue [ rdf:value ?latencyCost ] .
  ?capability sec:performanceMetric [ sec:MeanTimeToMitigate ?mitigateTime ] .
}
\end{lstlisting}
\end{figure}

The graph engine processes the query and returns candidates \texttt{ex:DpiCapability} and \texttt{ex:ThrottlingCapability}, alongside their native costs.

\begin{enumerate}
    \setcounter{enumi}{2}
    \item \textbf{Constraint Solving:} The agent evaluates candidates against the conditions defined in \texttt{ex:DdosMitigationIntent}:
    \begin{itemize}
        \raggedright
        \item \textbf{Evaluation of Candidate A (\texttt{ex:DpiCapability}):} Mitigation speed is optimal ($10\text{ ms} \leq 50\text{ ms}$), but its resource impact value on \texttt{ex:NetLatency} is $35\text{ ms}$. This violates the configured ceiling,\texttt{ex:MaxLatencyCeiling}, ($35\text{ ms} > 20\text{ ms}$), so Candidate A is discarded.
        \item \textbf{Evaluation of Candidate B (\texttt{ex:ThrottlingCapability}):} Mitigation speed is $30\text{ ms} \leq 50\text{ ms}$. Its latency resource impact value is $15\text{ ms}$, which fully complies with the operational performance ceiling ($15\text{ ms} \leq 20\text{ ms}$).
    \end{itemize}
    \item \textbf{Enforcement:} The IMF selects Candidate B, invokes the control protocol to initiate dynamic edge request throttling, and safely neutralizes the signaling storm while maintaining SLA integrity. 
\end{enumerate}




\section{Open Challenges and Future Direction}

A primary challenge in deploying model-driven architectures to real-time production systems is the latency footprint of graph query parsing. While our choice of building TR292I strictly around an RDFS hierarchy explicitly avoids the computational heavy lifting and state explosion bottlenecks of traditional OWL Web Ontology Language description logic reasoners, processing graph pattern-matching execution loops over large triple stores remains demanding \cite{mehmood2024securing}. 

In a dense, next-generation 6G ecosystem featuring millions of ephemeral, fluid network components and slicing interfaces, complex graph queries evaluation across large semantic graphs could take several milliseconds. This delay could exceed the short mitigation windows required to counter sophisticated control-plane saturation campaigns. Future work must explore structural performance optimizations, including decentralized subgraph partitioning, incremental graph update indexing, and deploying hardware-accelerated graph database engines directly at the cell site or network edge.

In production, concurrent intents from separate administrative domains inevitably conflict on shared infrastructure. For instance, a business-plane intent prioritizing channel throughput may conflict with a security-plane intent invoking request throttling. Concurrently, an active security-plane intent reacting to an ongoing signaling anomaly may invoke deep traffic monitoring filters or request throttling actions that directly reduce total channel capacity. 

Because TR292I models performance impact values explicitly, it provides the necessary semantic infrastructure to detect these runtime cross-domain policy collisions. Developing automated, state-aware conflict-resolution strategies and hierarchical optimization matrices remains a critical path for future research.

\section{Conclusion}
This work introduces the design, architectural integration, and standardization of the TM Forum TR292I Security Ontology v4.0.0, establishing a robust framework for declarative, self-protecting network operations. By extending the TM Forum ICM with an RDFS compliant vocabulary, the proposed architecture enables network orchestrators to map complex attack surfaces and execute mitigation strategies using an intent-driven paradigm. 

The integration of a mathematical multi-constraint optimization engine guarantees that security enforcement functions neutralize active threat vectors without compromising active customer SLAs. In this work we provided a structural validation via a formal semantic walkthrough of a DDoS mitigation sequence on a disaggregated gNB slice, as one example of a threat. This validation confirms that the framework effectively arbitrates runtime policy conflicts without human intervention. 

Ultimately, this standardized model provides the semantic interoperability, structured bounds, and architectural consistency necessary to secure fluid topologies within 5G-Advanced and future 6G networks.

\begin{small}
\bibliographystyle{IEEEtran}
\bibliography{references}
\end{small}

\end{document}